\documentclass[preprints,article,accept,moreauthors,pdftex]{Definitions/mdpi}
\usepackage{amsmath,amssymb,amsfonts}
\usepackage[cal=cm]{mathalfa}
\usepackage{siunitx}
\usepackage{multirow}
\firstpage{1}
\pubvolume{1}
\issuenum{1}
\articlenumber{0}
\pubyear{2021}
\copyrightyear{2020}
\datereceived{27 January 2022} 
\dateaccepted{28 February 2022} 
\datepublished{} 
\hreflink{https://doi.org/} 
\newcommand{\be}{\begin{equation}}
\newcommand{\ee}{\end{equation}}
\newcommand{\daa}{\Delta\alpha/\alpha}
\newcommand{\dll}{\Delta\lambda/\lambda}

\pdfoutput=1

\Title{Avoiding bias in measurements of fundamental constants from high resolution quasar spectra\footnote{This article is a contribution for the AlteCosmoFun'21 conference volume, 6-10 September 2022, Szczecin, Poland, organised by the Szczecin Cosmology Group.}}

\TitleCitation{Avoiding bias in measurements of fundamental constants from high resolution quasar spectra}

\Author{John K. Webb $^{1}$\orcidA{}, Chung-Chi Lee $^{1}$\orcidB{}, Dinko Milaković $^{2,3,4}$\orcidC{}}

\AuthorNames{John K. Webb, Chung-Chi Lee, Dinko Milaković}

\AuthorCitation{Webb, John K.; Lee, C-C.; Milaković, D.}

\address{%
$^{1}$ \quad Clare Hall, University of Cambridge, Herschel Rd, Cambridge CB3 9AL\\
$^{2}$ \quad Institute for Fundamental Physics of the Universe, Via Beirut, 2, 34151 Grignano, Italy\\
$^{3}$ \quad INAF -- Osservatorio Astronomico di Trieste, via Tiepolo 11, 34131, Trieste, Italy\\
$^{4}$ \quad INFN, Sezione di Trieste, Via Bonomea 265, 34136 Trieste, Italy\\
}

\abstract{Recent advances in spectroscopic instrumentation and calibration methods dramatically improve the quality of quasar spectra. Supercomputer calculations show that, at high spectral resolution, procedures used in some previous analyses of spacetime variations of fundamental constants are likely to generate spurious measurements, biased systematically towards a null result. Developments in analysis methods are also summarised and a prescription given for the analysis of new and forthcoming data.}


\begin{document}

\section{Introduction}

High resolution spectra of distant quasars reveal numerous narrow absorption lines caused by gaseous components of galaxies intersecting the Earth--quasar sightline. The large number of atomic species and transitions detected allow precise measurements of the fine structure constant $\alpha_{SI} = e^2/4\pi\epsilon_0\hbar c$ over cosmological distances, where $e$ is the electron charge, $\epsilon_0$ the vacuum permittivity, $\hbar$ the reduced Planck constant, and $c$ the speed of light. The dimensionless $\alpha$ is the ratio of the speed of an electron in the lowest energy orbit of the Bohr-Sommerfeld atom to the speed of light, and hence connects quantum mechanics (through $\hbar$) with electromagnetism (through the remaining quantities).

The 1999 invention of the {\it Many Multiplet method} (MM) \cite{Dzuba1999, Webb1999} created an order of magnitude precision gain over previous methods in searches for spacetime variations of $\alpha$. The previously used method had been the {\it Alkali Doublet method} (AD), in which measurements were made relative to the same ground state, by-passing the invaluable sensitivity of the ground state to any change in $\alpha$. Further, the excited levels in the $^2P_{3/2}$--$^2P_{1/2}$ AD structure itself generally have lower sensitivity compared to the commonly observed singly ionised multiplets of Fe, Zn, Cr, and others. In contrast, the MM method takes into account ground-state shifts. Also, excited state relativistic corrections in multiplets can be large; $s$--$p$ and $s$--$d$ transitions for example may even be of the opposite sign. Any real change in $\alpha$ therefore generate a unique pattern of observed wavelength shifts that is not degenerate with a simple cosmological redshift. The MM method thus produces sensitive results when applied to multiplets of the same atomic species, or to species having widely differing atomic masses.

New and forthcoming scientific facilities\footnote{In particular, the Echelle SPectrograph for Rocky Exoplanets and Stable Spectroscopic Observations (ESPRESSO) on the European Southern Observatory's Very Large Telescope (VLT) \cite{espresso2021} and the High Resolution Echelle Spectograph (HIRES) on the forthcoming Extremely Large Telescope (ELT) e.g. \cite{Marconi2016, ELT2018}} will intensify searches for spacetime variations of fundamental constants. As the data quality and quantity increase, it becomes all-important to ensure that analysis techniques produce fully unbiased and optimal estimates. Here we summarise recent methodological advances facilitating these things and to scrutinise assumptions made and procedures used in previously published analyses that can produce bias. The remainder of this paper is composed as follows: Sections \ref{sec:wavelengths} and \ref{sec:vpfit} summarise several recent advances in the varying $\alpha$ field. Section \ref{sec:blinding} presents some new calculations showing that ``blinding'' methods, as used in some published measurements, generate measurement bias. Section \ref{sec:howto} gives a pr\'ecis of the do's and don'ts when analysing high quality absorption spectra.

\section{Wavelength calibration} \label{sec:wavelengths}
Measuring $\alpha$ requires the wavelength scale of the astronomical spectrum to be established with high fidelity. For example, a change of $\daa \equiv (\alpha_z - \alpha_0)/\alpha_0  =1\times10^{-6}$ gives rise to a relative shift in the wavelength of Fe\,{\sc ii} 2383 of approximately $\dll=1\times10^{-7}$. For echelle spectrographs such as UVES on the VLT, the standard wavelength reference is established by imaging the spectrum of a ThAr arc lamp. A comparison between ThAr wavelength calibrated data and the solar spectrum \cite[as first done by][]{Molaro2008}, revealed the presence of long-range wavelength scale distortions with amplitudes as large as $\dll=1\times10^{-5}$. It was initially thought that distortions this large could potentially spoil fundamental constant measurements \cite{Rahmani2013,Whitmore2015}, although subsequent analyses showed that such distortions can be modelled and associated uncertainties on $\daa$ allowed for (see \cite{Dumont2017} and Appendix B9 of \cite{WebbVPFIT2021}). 

The new generation of astronomical spectrographs aims to avoid the long-range distortions seen in some ThAr calibrated spectra altogether. Laser Frequency Comb technology \citep[LFC,][]{Haensch2006,Steinmetz2008,Haensch2013} or Fabry-Per{\'o}t etalons (FP) combined with ThAr can provide wavelength calibration with accuracy of the order $\dll=1\times10^{-8}$ \cite{Milakovic2020,Probst2020,Schmidt2021}. Both of these advanced calibration sources are installed on the High Accuracy Radial-velocity Planet Searcher \cite[HARPS,][]{Mayor2003} and ESPRESSO. One current difficulty is that none of the currently available astronomical LFCs provide wavelength calibration below 5000 {\AA}. Wavelengths below this cut-off are currently calibrated by simultaneously imaging both ThAr and FP spectra. The ThAr + FP combination provides lower accuracy than LFCs. Since the spectral region below 5000 {\AA} is generally very important for $\daa$ measurements and since, this important problem is yet to be solved, raising significant problems not only for varying constant measurements but also for redshift drift projects.

Another important concern that must be addressed in future high-precision spectroscopy is that using an LFC or ThAr+FP may not be sufficient to remove all systematic effects related to wavelength calibration. A comparison of two independent LFCs used simultaneously on HARPS revealed an unexpected offset in the zero-points of their wavelength calibrations of $\dll=1.5\times10^{-9}$ \cite{Probst2020,Milakovic2020}. Although this is a very small effect, combining observations calibrated using either two different LFCs or a single LFC that has been modified in some way between two observations should be carefully performed.

\section{Absorption profile modelling} \label{sec:vpfit}

\subsection{VPFIT and AI-VPFIT}

{\sc vpfit}\footnote{\url{https://people.ast.cam.ac.uk/~rfc/}} is a non-linear least squares code for modelling high resolution absorption spectra that has been developed over a number of years \cite{ascl:VPFIT2014, web:VPFIT} and it forms the core of our procedures. The theoretical background on which {\sc vpfit} is based, plus some recent enhancements, are described in \cite{WebbVPFIT2021, Lee2021Addendum}. Throughout {\sc vpfit}'s development, considerable effort has gone into ensuring high precision internal calculations. It has a comprehensive online user guide, updated frequently\footnote{The current {\sc vpfit} user guide is available at \url{https://www.overleaf.com/read/vbxkcfnfgksr}}. The ever-increasing quality of high resolution spectroscopic data requires extremely careful treatment of every aspect of profile calculation and fitting. Some of the studies carried out in this context are:
\begin{enumerate}
    \item The Voigt profile 2-level atom approximation is good enough, even at very high spectral signal to noise, for non-damped column densities i.e. $\log N \leq 20$ \cite{Lee2020KHT}.
    \item Voigt function $H(a,u)$ look-up tables must be sufficiently high resolution in the relevant parameter ($u$) to render non-linear effects negligible. Interpolation within those tables must also be sufficiently precise \cite{WebbVPFIT2021}.
    \item It is most important to allow for non-linearity in the Voigt profile shape by computing model Voigt profiles in extremely fine bins \cite{ascl:VPFIT2014}.
    \item As far as possible, any blends/interlopers must be allowed for, whether they arise in identified species or not. Failure to do can significantly increase the measurement error and bias individual measurements \cite{Lee2020AI-VPFIT}.
    \item Fitting region selection is important. If line wings/continuum regions are truncated, best-fit models for $\daa$ exhibit an unnecessarily large scatter \cite{Wilczynska2015}.
    \item Voigt function derivatives must be accurate \cite{WebbVPFIT2021, Lee2021Addendum}. The Hessian is derived from the derivatives of $H(a,u)$, the inverse of which provides parameter uncertainties. The Hessian, of course, determines parameter search directions.
    \item How to select a ``final'' absorption system model, given non-uniqueness and alternative information criteria? \cite{Lee2021}.
    \item Absorption line broadening: models must include temperature as a free parameter and should not be assumed to be {\it turbulent} \cite{Milakovic2021, Noterdaeme2021}.
    \item Using the correct instrumental profile for model calculations is important \cite{Milakovic2020}.
    \item All spectral regions used in the measurement must be carefully checked for potential contaminating atmospheric features\footnote{\url{http://www.eso.org/observing/etc/bin/gen/form?INS.MODE=swspectr+INS.NAME=SKYCALC}} \cite{web:skycalc}.
\end{enumerate}

Using {\sc vpfit} involves human decision making and the final results obtained can depend on that human input \cite{Lee2021}. Whilst this has negligible consequences for many applications, measurements of fundamental constants at high redshift push the limits of the data and are more susceptible to small systematic errors or biases than other less challenging measurements. Measurements of $\daa$ thus motivate full automation in which all human input is avoided and all potential bias removed. Artificial Intelligence methods were first applied to this problem in \citep{Bainbridge2017, gvpfit2017} ({\sc gvpfit}) and significantly extended in \cite{Lee2020AI-VPFIT} ({\sc ai-vpfit}). The advantages in avoiding human decision making are achieving objectivity, reproducibility, and the ability to explore multiple models to the same absorption system. These advantages turn out to be crucial for measurements of fundamental constants. The calculations reported in the following sections make use of both {\sc vpfit} and {\sc ai-vpfit}.

\subsection{Information criterion or $\chi^2_{\nu}$ to select models?}

In selecting a single best-fit model for an absorption complex, many previous measurements have made use of a simple normalised $\chi^2$ approach, accepting a model once $\chi^2_{\nu}$ is ``sufficiently'' close to unity,
\be
\chi^2_{\nu} = \frac{1}{\nu} \sum_{i=1}^{n_d} \left( \frac{d_i - f_i}{\sigma^2_i} \right)^2
\label{eq:chisqn}
\ee
where $d_i$ is the spectral data array, $f_i$ is the model, $\sigma_i$ is the spectral error array, $n_d$ is the number of data points, and the number of degrees of freedom $\nu=n_d - n_p$, where $n_p$ is the total number of free parameters in the model. Alternatively, one can use an {\it information criterion} (IC) to select models, the general form of which is
\be
\textrm{IC} = \chi^2 + \mathcal{P}(n_p,n_d)
\label{eq:IC}
\ee
where $\chi^2 = \nu\chi^2_{\nu}$ and $\mathcal{P}(n_p,n_d)$ is a penalty factor that increases with increasing number of model parameters. An IC applies a ``principle of parsimony'', balancing parameter variance with model bias. The application of ICs in astrophysics has been discussed by \cite{Liddle2004, Liddle2007} and a comprehensive treatise is given in \cite{Burnham2002}. Problems associated with noise characteristics in calculating ICs are discussed in \cite{Rossi2020}. Employing an IC allows an optimal number of model parameters to be identified in an objective and reproducible way \cite{Webb2021}.

Using an asymptotic $\chi^2_{\nu}$ as a means of deciding how many model parameters to use requires the user to choose a maximum acceptable value for $\chi^2_{\nu}$. This in itself is not particularly disadvantageous because one can formalise the problem through the relationship between $\chi^2_{\nu}$ and its probability distribution, hence defining an acceptance probability rather than a numerical value of $\chi^2_{\nu}$ for model selection. However, in practice it is generally difficult to obtain $\sigma_i$ accurately, so $\chi^2$ itself is only approximate\footnote{The difficulties in accurately estimating the spectral error array are well known and have been discussed in the section titled {\it Modifying the error arrays} in the {\sc rdgen} user guide \cite{web:VPFIT}.}. 

For the reasons outlined above, in the context of $\daa$ at least, it is preferable to use an IC for model selection rather than $\chi^2_{\nu}$. However, using an IC raises a question: what is the optimal form of the penalty term $\mathcal{P}(n_p,n_d)$ in Equation \eqref{eq:IC}? This point has been studied in \cite{Webb2021}, where three ICs are compared : the corrected Akaike Information Criterion AICc, the Bayesian Information Criterion BIC, and the Spectral Information Criterion SpIC. The latter is a new IC designed specifically for spectroscopy. All three ICs (there are others) perform in slightly different ways; AICc tends to over-fit the data, allowing too many free parameters, whilst the converse is true for BIC. SpIC appears to fall in the Goldilocks zone.

\subsection{Contributions to the $\daa$ error budget}

Fully understanding all potential contributions to the error budget, random and systematic, is of course a crucial aspect of assessing the reality of any deviation of $\daa$ from zero. Uncertainties associated with varying $\alpha$ measurements have been discussed in numerous papers e.g. \cite{Murphy2001, King2012}. Since our understanding of uncertainties has significantly improved since those studies, we list here possible sources of error and their attribution: \vspace{0.07in} \\ 
{\it Inherent uncertainties:}
\begin{enumerate}
\item[1.] Statistical error i.e. VPFIT covariance matrix error.
\item[2.] If turbulent broadening is used to model the system and if the true intrinsic broadening is {\it compound} (see Eq. \eqref{eq:btot}), a systematic error is introduced.
\item[3.] Absorption system model non-uniqueness error.
\item[4.] Continuum estimate error.
\end{enumerate}
{\it Errors from astrophysical factors:}
\begin{enumerate}
\item[5.] Isotopic relative abundances.
\end{enumerate}
{\it Errors associated with theoretical and experimental uncertainties:}
\begin{enumerate}
\item[6.] Q coefficient uncertainties (these bias towards $\daa=0$).
\item[7.] Oscillator strength uncertainties.
\item [8.] Laboratory wavelength uncertainties.
\end{enumerate}
{\it Errors associated with data extraction or instrumental factors:}
\begin{enumerate}
\item[9.] Wavelength calibration error (for pre-LFC/FP data).
\item[10.] Bad pixels.
\item[11.] Flat-fielding errors.
\item[12.] Weak cosmic rays removal.
\item[13.] Significant point-spread function variations across the detector.
\end{enumerate}

\subsection{Future measurements require Monte Carlo AI} \label{sec:MC}

The recent application of AI methods ({\sc ai-vpfit}) provides full automation of modelling quasar absorption systems \cite{gvpfit2017, Lee2020AI-VPFIT}. Automation has allowed us to explore in detail how final best-fit models depend on the construction sequence and hence the extent to which model non-uniqueness contributes to the total $\daa$ error budget.  Preliminary studies of this sort have been made recently and so far indicate that indeed $\daa$ measurements are impacted by the sequence in which models are developed. The inference is that, for any particular absorption system, it is difficult to properly assess the overall uncertainty on $\daa$ {\it unless} multiple models are produced, each one constructed differently (emulating the different approaches that would be taken by different human modellers). The contribution of non-uniqueness to the overall $\daa$ uncertainty budget can only be determined on a case by case basis because we now know that the degree of non-uniqueness differs from one absorption system to another \cite{Lee2021}. In other words, it is necessary to form many models of each particular absorption system in order to quantify the intrinsic non-uniqueness behaviour and hence determine the overall uncertainty on $\daa$.

\section{Spectral simulations and distortion-blinding} \label{sec:blinding}

The methods used to measure $\daa$ must be entirely free of any bias. To that end, a ``blinding'' method has been used in several published measurements \cite{Evans2014, Murphy2017, Kotus2017, Murphy2021}, employing the following steps. During the initial data reduction stages, long-range and intra-order distortions of the wavelength scale are applied to individual exposures. The individual distortions are designed to leave a non-zero $\daa$ in the final co-added spectrum. {\sc vpfit} is then used on that distorted spectrum, fixing $\daa$ to be zero. Once the final model has been obtained in this way, one final tweak of the model is carried out by fitting a non-distorted version of the spectrum, allowing the existing model parameters plus $\daa$ to vary freely. However, fixing $\daa=0$ corresponds to requiring that all rest-frame wavelengths involved in the initial absorption system modelling are precisely those measured in terrestrial laboratories. If the true value of $\daa$ is zero, the method just described should indeed be unbiased. However, if the true $\daa$ is {\it not} zero (and it is explicitly non-zero in the ``blinded'' data of \cite{Evans2014, Murphy2017, Kotus2017, Murphy2021}), forcing $\daa$ to be initially zero necessarily produces a flawed model. The final step of ``switching on'' $\daa$ as a free parameter may or may not subsequently be able to correct that flawed model. If it does not, the result is a systematically biased measurement. Notwithstanding the previous comments, it should be noted that the spectral resolution of the simulations in the analyses in this paper are far higher than in e.g. \cite{Evans2014, Murphy2017, Kotus2017}. At lower spectral resolution, fewer components are detected. This could mean that the level of bias is diluted. Further calculations are needed to answer this point. Irrespective of this, what is instead needed is a method that is fully unbiased, {\it no matter what} the true value of $\daa$ is. We next describe some preliminary simple spectral simulations to illustrate the concerns just expressed. To distinguish the sort of blinding described above from other potential blinding methods, we will refer to the method described above as ``distortion-blinding''.

\subsection{Preliminary illustration of how distortion-blinding + turbulent line broadening creates bias} \label{sec:preliminary}

\begin{table}
{\small
\begin{tabular}{|c|c|c|c|c|c|c|}
\hline
 & Species & $\log N$ & $b_{turb}$ (km/s) & $T$ ($10^4$ K) & $v$ (km/s) & $\Delta \alpha / \alpha$ ($10^{-5}$) \\
\hline
Generating & Mg\,{\sc ii} & 12.500        & 3.87 & 5.31 & 0.000 & 10.0 \\
model      & Fe\,{\sc ii} & 12.000 &      &      &      &      \\ \hline
Fitting    & Mg\,{\sc ii} & 12.499 (0.002) & 3.70 & 5.53 & 0.000 & 10.1 \\
(compound) & Fe\,{\sc ii} & 12.000 (0.007) &(0.27)&(0.34)&(0.028)&(0.4) \\ \hline
           & Mg\,{\sc ii} & 12.493 (0.017) & 7.02 & - &-0.279 &   \\
Fitting    & Fe\,{\sc ii} & -              &(0.05)& &(0.084)& - \\ \cline{2-6}
(turbulent)& Mg\,{\sc ii} & 10.682 (1.099) & 5.44 & - &-2.152 &  \\
           & Fe\,{\sc ii} & 12.001 (0.007) &(0.10)& &(0.070)&  \\ \hline
\end{tabular}}
\caption{Results obtained using {\sc vpfit} on a single component simulation. See Section \ref{sec:preliminary} for details. $v = (z-z_0)c/(1+z_0)$, where $z_0 = 1.1469691$ and $c$ is the speed of light. In the lower panel, one of the Fe\,{\sc ii} components fell below the {\sc vpfit} detection limit, indicated in the $\log N$ column by ``-''. The negative velocities ($v$) for the turbulent case are a consequence of forcing $\daa=0$ (the true value is non-zero). \label{tab:single}}
\end{table}

\begin{figure*}
\centering
\includegraphics[width=0.98\linewidth]{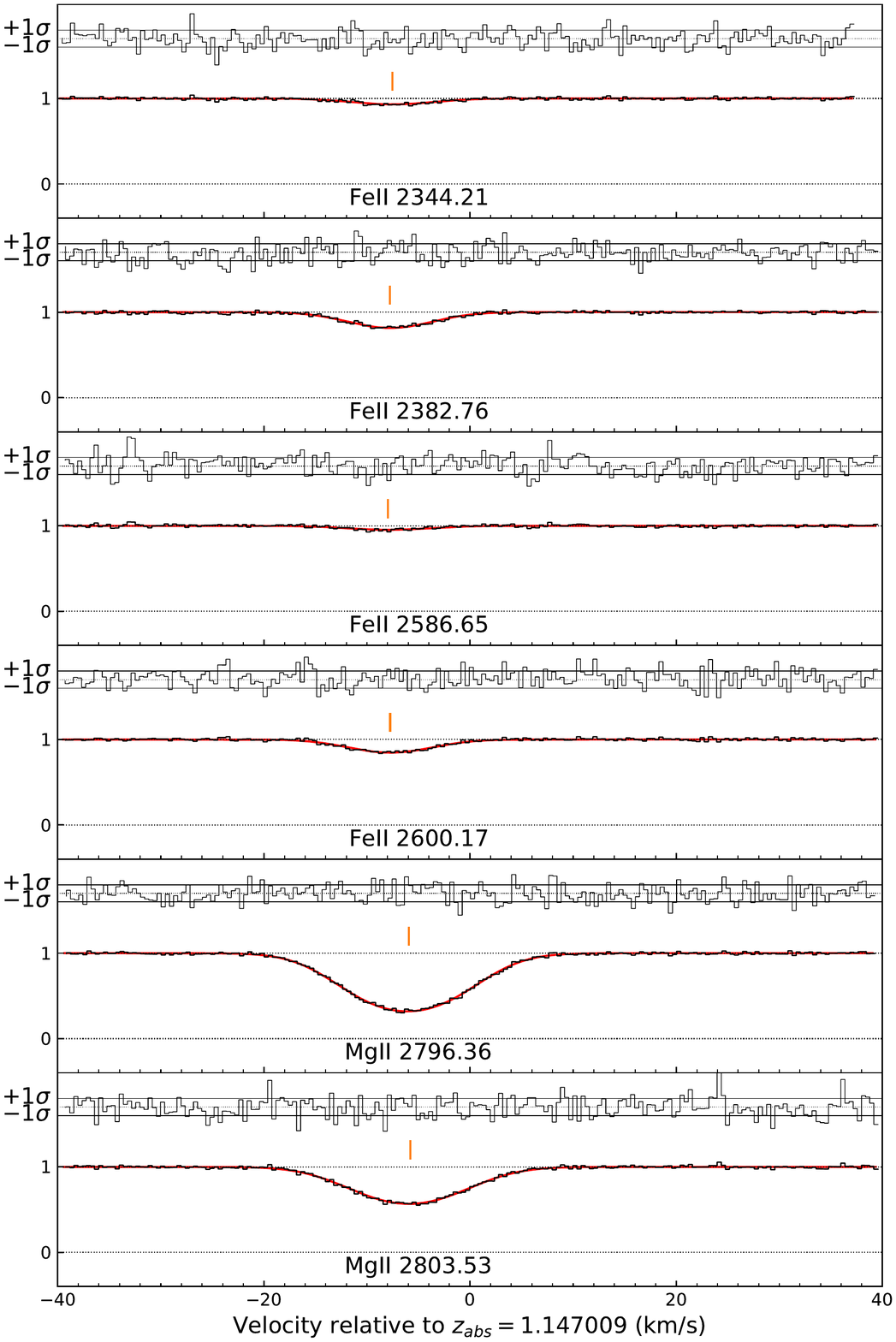}
\caption{The data (black histogram) in this plot corresponds to the upper panel labelled ``Generating model'' in Table \ref{tab:single}. The model (continuous red curve) corresponds to the middle panel labelled ``Fitting (compound)''. See Section \ref{sec:preliminary}.
\label{fig:exaggerated_compound}}
\end{figure*}

\begin{figure*}
\centering
\includegraphics[width=0.98\linewidth]{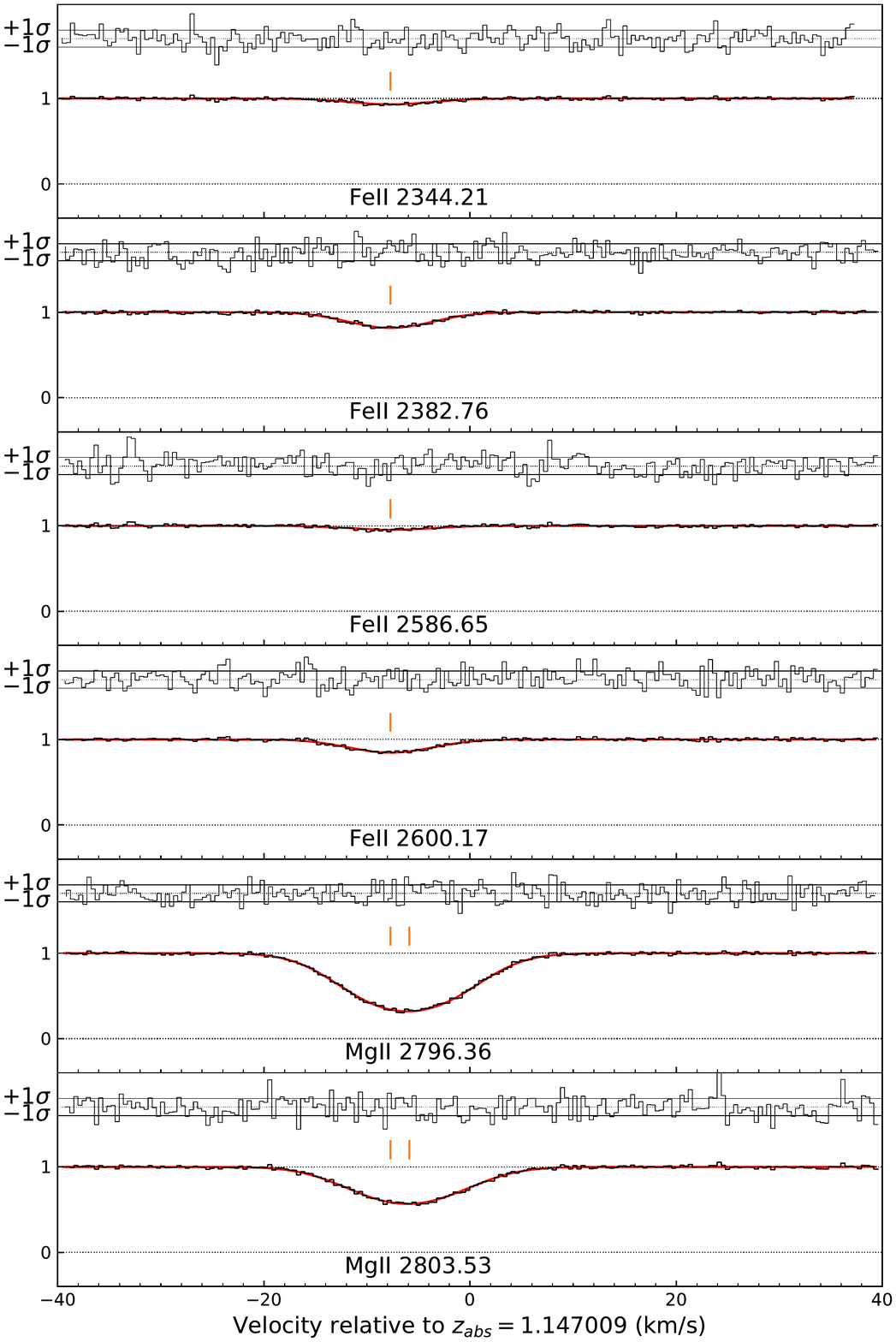}
\caption{The data (black histogram) in this plot corresponds to the upper panel labelled ``Generating model'' in Table \ref{tab:single}. The model (continuous red curve) corresponds to the middle panel labelled ``Fitting (turbulent)''. See Section \ref{sec:preliminary}.
\label{fig:exaggerated_turbulent}}
\end{figure*}

A simulated spectrum of a single component absorption system is created, with seven transitions: Mg\,{\sc ii} 2796, 2803, Fe\,{\sc ii} 2344, 2374, 2383, 2586, 2600 {\AA}. A spectral resolution of $2.10$ km/s FWHM is used for the three transitions having $\lambda_{obs}<5120$ {\AA} and $2.05$ km/s FWHM for the four transitions having $\lambda_{obs} > 5540${\AA}. The pixel size and signal to noise per pixel were 0.4 km/s and 75. The spectral parameters correspond to those of the ESPRESSO spectrum in \cite{Murphy2021}. The absorption line parameter values used are similar to one component from the $z_{abs}=1.15$ system towards the quasar HE0515$-$4414. The upper panel of Table \ref{tab:single} (labelled ``Generating model'') shows the actual absorption line parameters used to generate the simulated spectrum. For the purposes of this preliminary illustration, we use an extreme value of $\daa=10^{-4}$, so that line shifts caused by the non-zero $\daa$ are easily visible in the plotted data. A subset of the simulated spectral lines is illustrated in Figure \ref{fig:exaggerated_compound}. As Table \ref{tab:single} shows, the simulated spectrum is generated using compound broadening i.e. line broadening comes from both turbulent and thermal contributions,
\be
b_{obs}^2 = b_{\textrm{turb}}^2 + \frac{2kT}{m}
\label{eq:btot}
\ee
Having created the data, for the purposes of this illustration we first fit all lines simultaneously with {\sc vpfit}, using compound line broadening. $\daa$ is a free parameter throughout. No distortion-blinding is used. The second panel of Table \ref{tab:single} (labelled ``Fitting (compound)'') illustrates the measured parameter values and their uncertainties, showing, as expected, that only a single absorption component is required and that all parameters returned are consistent with the input values.

We now reach the point of this illustration. The seven transitions are again fitted simultaneously using {\sc vpfit}, but this time with two important differences: the data are fitted using a fixed $\daa=0$ and using turbulent line broadening, i.e. taking the limiting case of $T=0$. The numerical results are given in the third (lower) panel of Table \ref{tab:single} (labelled ``Fitting (turbulent)'') and the resulting model is illustrated in Figure \ref{fig:exaggerated_turbulent}. This fit now requires {\it two} absorption components (one of the Fe\,{\sc ii} components is below the column density threshold so tick marks are not shown).  Figure \ref{fig:exaggerated_turbulent} and Table \ref{tab:single} reveal that adopting a turbulent model creates the necessity for an additional (fake) velocity component {\it and} because of the distortion-blinding, that fake velocity component is required to correspond perfectly with (redshifted) terrestrial laboratory wavelengths. 

The purpose of this simple simulation is purely to demonstrate how easy it is to derive a strongly biased result if distortion-blinding + turbulent broadening are used; the normalised residuals are plotted above each transition in Figure \ref{fig:exaggerated_turbulent}. The residuals show that the turbulent distortion-blinded fit forcing $\daa=0$  provides a good fit, even though it is far from zero in the data. The normalised $\chi^2$ for both fits quantify that: the non-blinded compound model produces an overall fit of $\chi_{\nu}^2=1.0250$ for 1394 degrees of freedom and AICc=1440.882. The distortion-blinded $\daa=0$ turbulent model produces $\chi_{\nu}^2=1.0291$ for 1393 degrees of freedom and AICc=1447.581. Both normalised $\chi^2$ values are acceptable. In Section \ref{sec:details} we examine this problem in more detail.

\subsection{Detailed calculations using {\sc ai-vpfit}} \label{sec:details}

Section \ref{sec:preliminary} illustrates that distortion-blinding + turbulent modelling has the capacity to significantly bias $\daa$ measurements towards zero. In this section we show that distortion-blinding + turbulent modelling does not merely have the capacity to bias, but that it inevitably does so. 

We first generate a simple synthetic spectrum. The absorption line parameters used were extracted from measurements (by us) of a real absorption system at $z_{abs}=1.15$ towards the publicly available ESPRESSO spectrum of the quasar HE0515$-$4414 \cite{Murphy2021}. A small section of the system was used, rather than the entire system, to make calculating time shorter and because a small section serves the required purpose. The spectral characteristics are the same as the simulated spectrum described in Section \ref{sec:preliminary}. The absorption line parameters used to generate the synthetic spectrum are given in Table \ref{tab:synth}. The synthetic spectrum was created using $\daa = 8.08 \times 10^{-6}$, a value measured by us, without applying any kind of blinding, using the full absorption system (i.e. not a small section). The same seven transitions were generated as above.

As Table \ref{tab:synth} shows, the line broadening used to generate the absorption simulation is compound i.e. each redshift has its own unique thermal and turbulent contribution. Many previous detailed analyses show that individual quasar absorption components exhibit compound and not turbulent broadening e.g. \cite{Noterdaeme2021}, so our synthetic spectrum emulates real data in this respect. The next stage of our analysis is to model the synthetic data using the same procedures that have been applied to real data in e.g. \cite{Murphy2021}. Models are fitted using turbulent line broadening and $\daa$ is fixed to zero i.e. we carry out the same distortion-blinding approach used in some previous works. We then apply {\sc ai-vpfit} \cite{Lee2020AI-VPFIT} multiple times, each time deriving an independently constructed best fit. AICc is used to select the best-fit model during each {\sc ai-vpfit} calculation. In each case, after the best-fit model is established, one final iteration using {\sc vpfit} is done, adding $\daa$ as an additional free parameter.

\begin{table}
\begin{tabular}{|l|l|l|l|l|}
\hline
log N(Mg\,{\sc ii}) & log N(Fe\,{\sc ii}) & $b_{turb}$ (km/s) & $T$ ($10^4$ K)  & $z_{abs}$ \\
\hline
12.66 & 12.20 &  1.49 &  1.64 & 1.1469691 \\
12.61 & 12.16 &  7.19 &  4.06 & 1.1469952 \\
\hline
\end{tabular}
\caption{Absorption line parameters used to generate the spectrum illustrated in Figure \ref{fig:synthetic_spectrum}. See Section \ref{sec:details}. \label{tab:synth}}
\end{table}

\begin{figure*}
\centering
\includegraphics[width=0.98\linewidth]{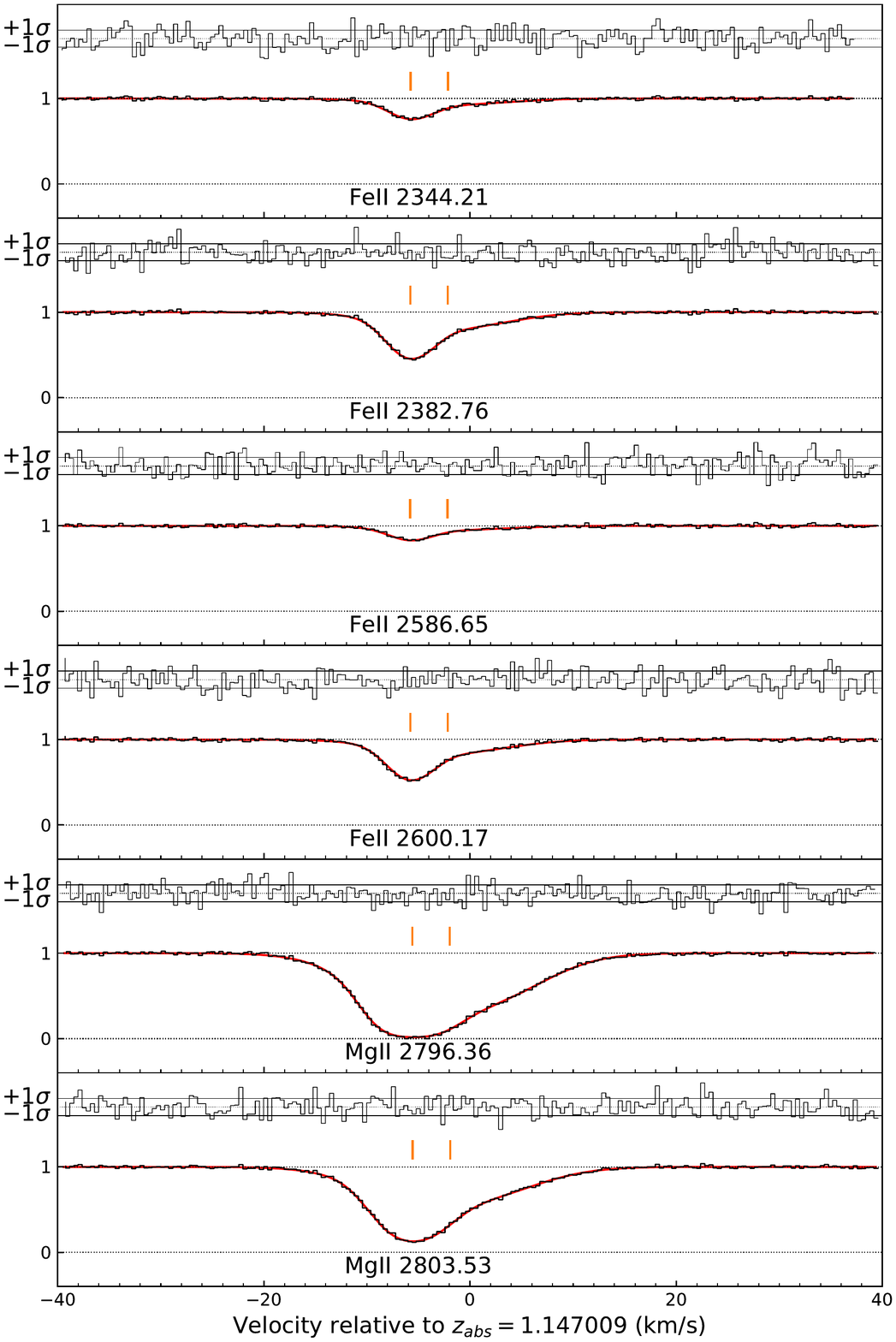}
\caption{Synthetic spectrum fitted with 4 sets of 100 {\sc ai-vpfit} models, as described in Section \ref{sec:details}. Only 6 transitions are shown (the excluded transition, Fe\,{\sc ii} 2374 {\AA}, is very weak). The parameter details are given in Table \ref{tab:synth}. The spectral characteristics are described in Section \ref{sec:preliminary}. The model (continuous red line) is a compound broadened model, randomly selected from our set of 100 fits. Vertical tick marks illustrate the best-fit component positions. \label{fig:synthetic_spectrum}}
\end{figure*}

\begin{figure}
\centering
\includegraphics[width=0.48\linewidth]{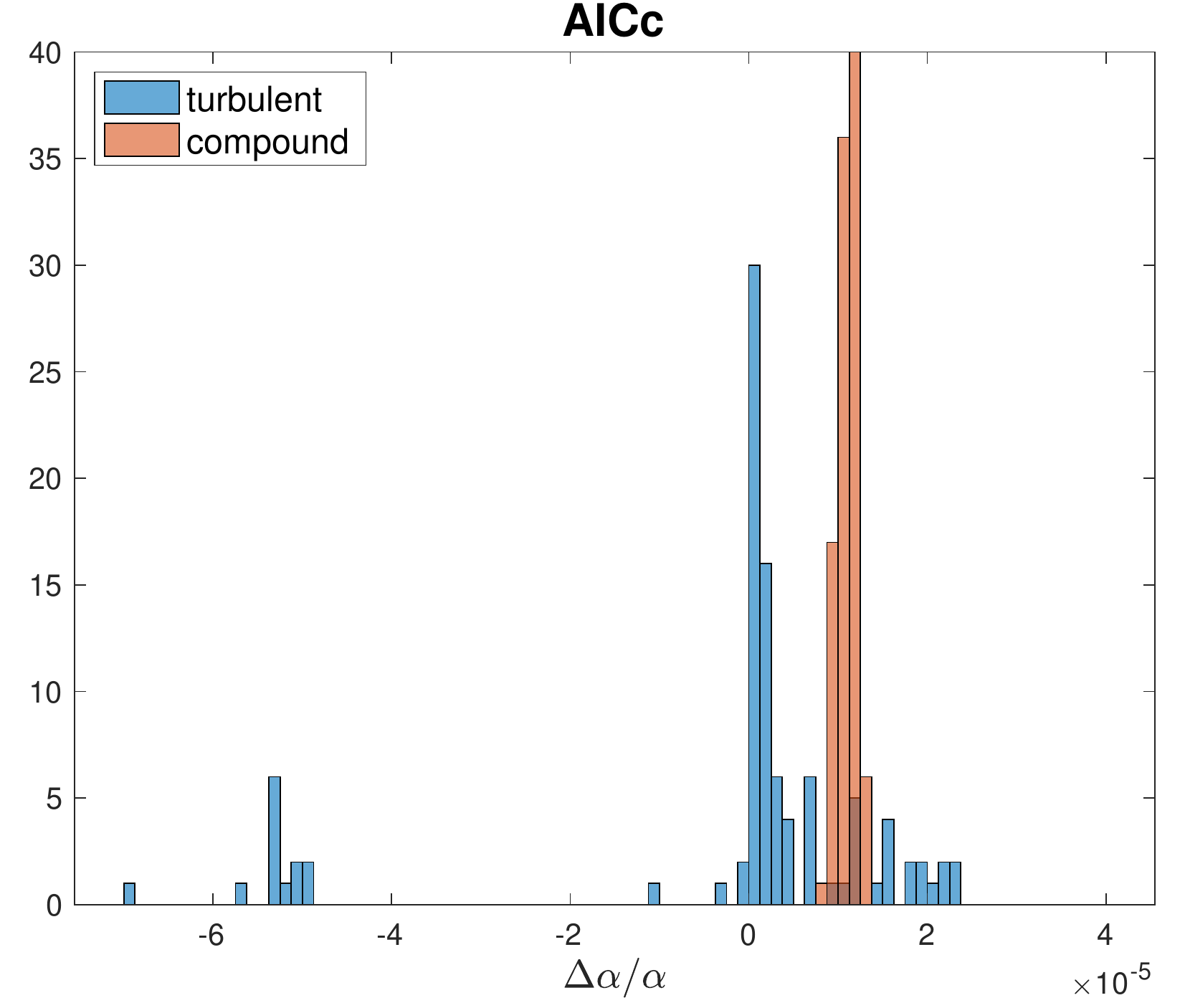}
\includegraphics[width=0.48\linewidth]{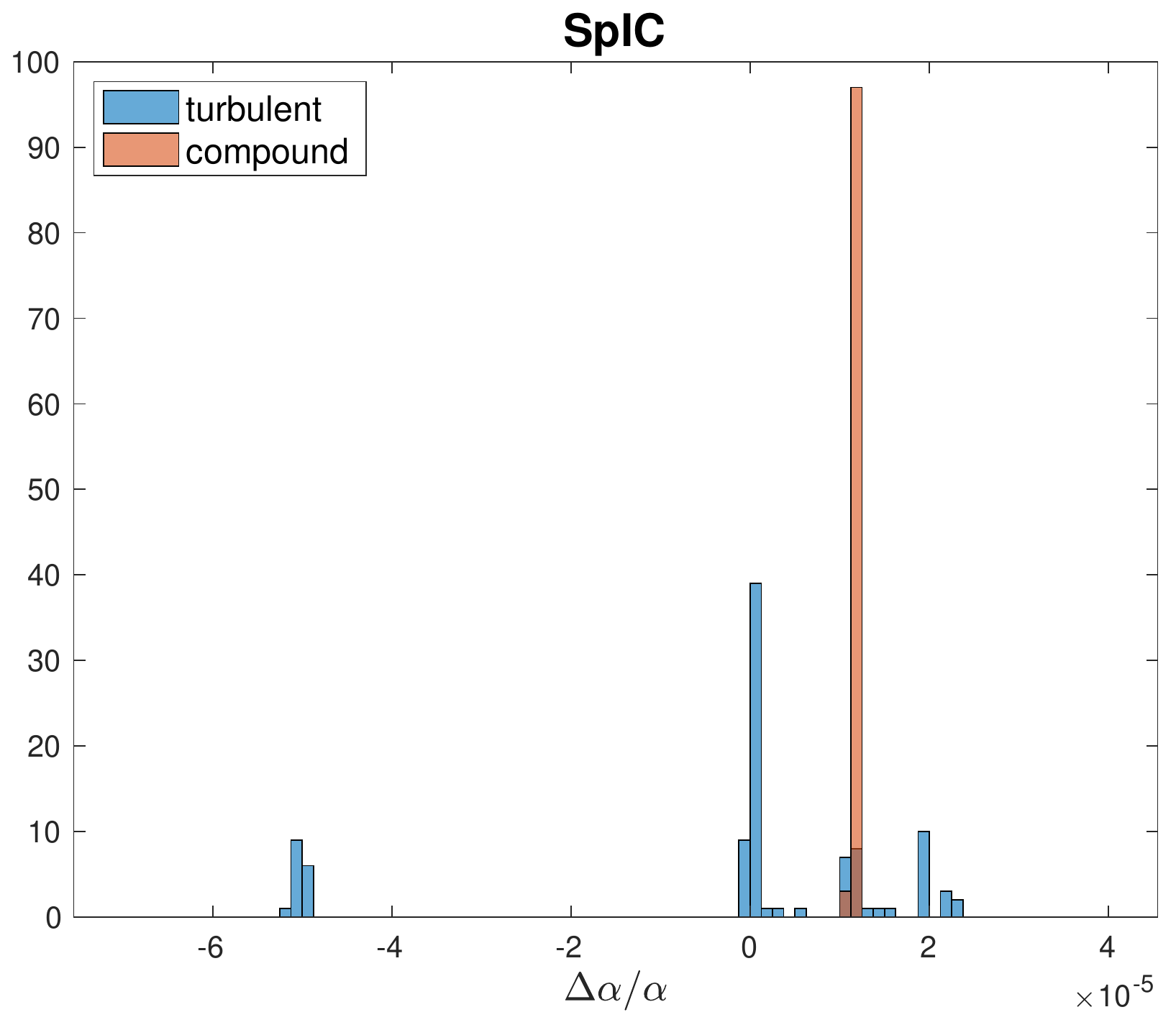}
\caption{Results from calculations simulating the impact of ``blinding'' on measurements of $\daa$. The red bars show the results from fitting compound broadened models without using distortion blinding. Those results are correct and unbiased. The blue bars show the results from fitting turbulent broadened models and using distortion blinding. The results are stongly biased and the correct $\daa$ is not recovered. See Section \ref{sec:details}. \label{fig:daa_blinding}}
\end{figure}

Figure \ref{fig:daa_blinding} gives histograms for four sets of 100 {\sc ai-vpfit} calculations (one for each combination of turbulent, compound, AICc, and SpIC models), revealing remarkable effects. The panel on the left demonstrates that using turbulent broadening with AICc generates a distribution that is clearly non-Gaussian, with severe bias towards $\daa=0$. No bias is seen for AICc/compound and the distribution looks well-behaved. The panel on the right shows results obtained using SpIC. In this case, significant bias is again seen if turbulent broadening is used, but not for compound broadening. The SpIC/compound $\daa$ results are more reliably determined than those for AICc/compound (compare the spreads of the compound points for AICc and SpIC results). The bin size in the histograms above is $1.25 \times 10^{-6}$ and the 1$\sigma$ statistical uncertainty for a typical compound broadened model with 2 components is $2.13 \times 10^{-6}$. Figure \ref{fig:daa_blinding} shows that the combination of distortion-blinding + turbulent broadening is particularly damaging to the final result.

\begin{figure}
\centering
\includegraphics[width=0.48\linewidth]{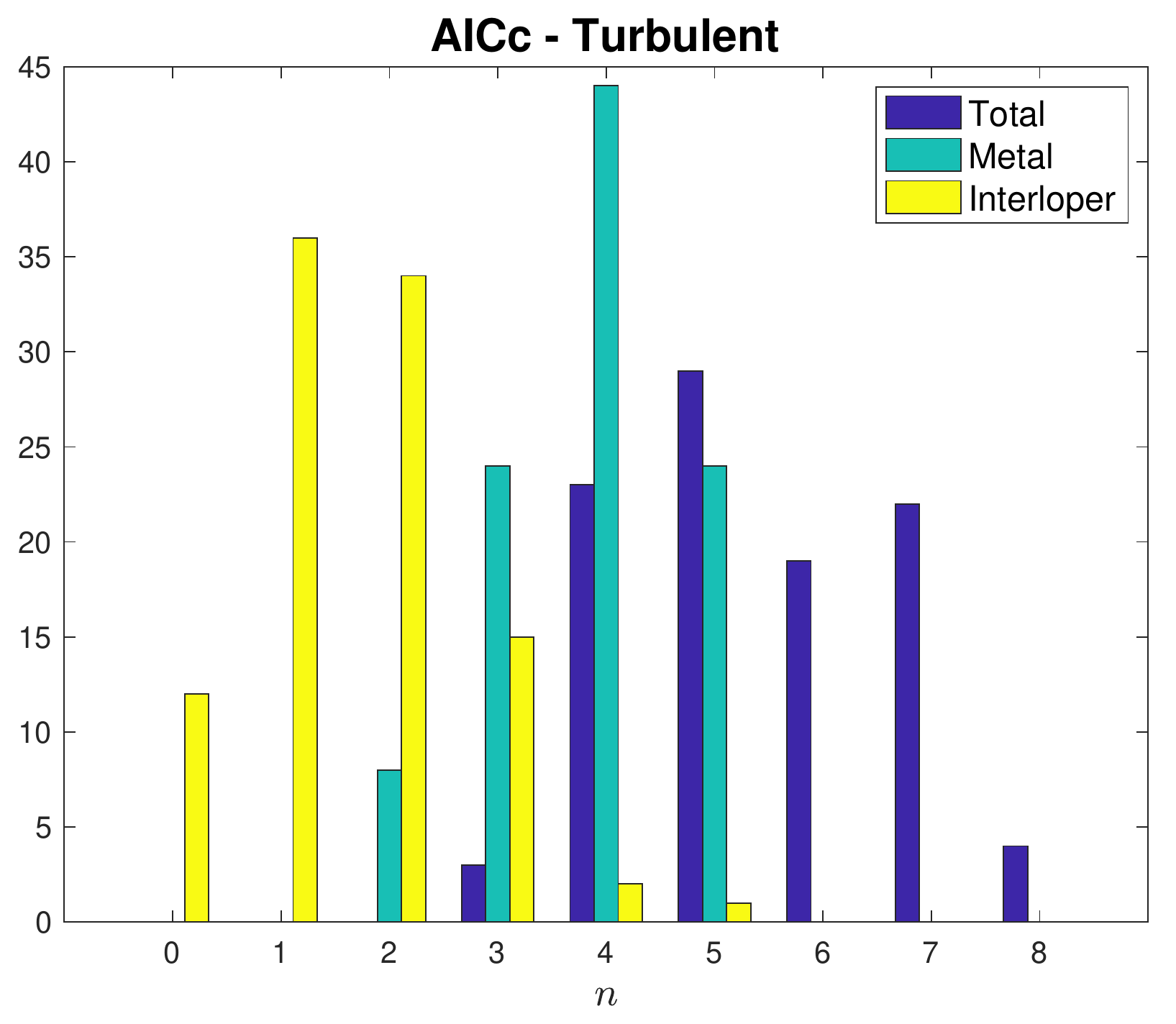}
\includegraphics[width=0.49\linewidth]{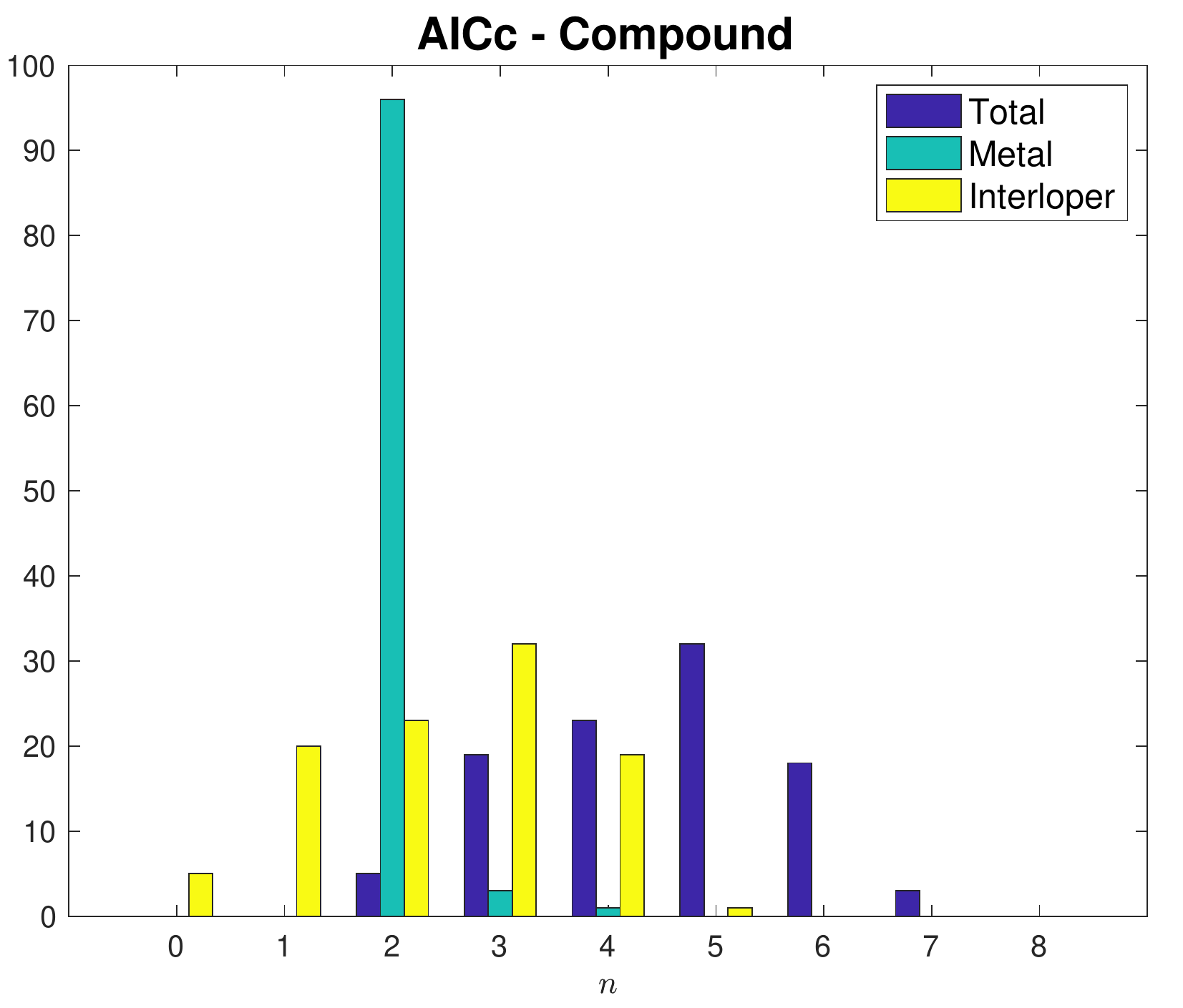} \\
\includegraphics[width=0.48\linewidth]{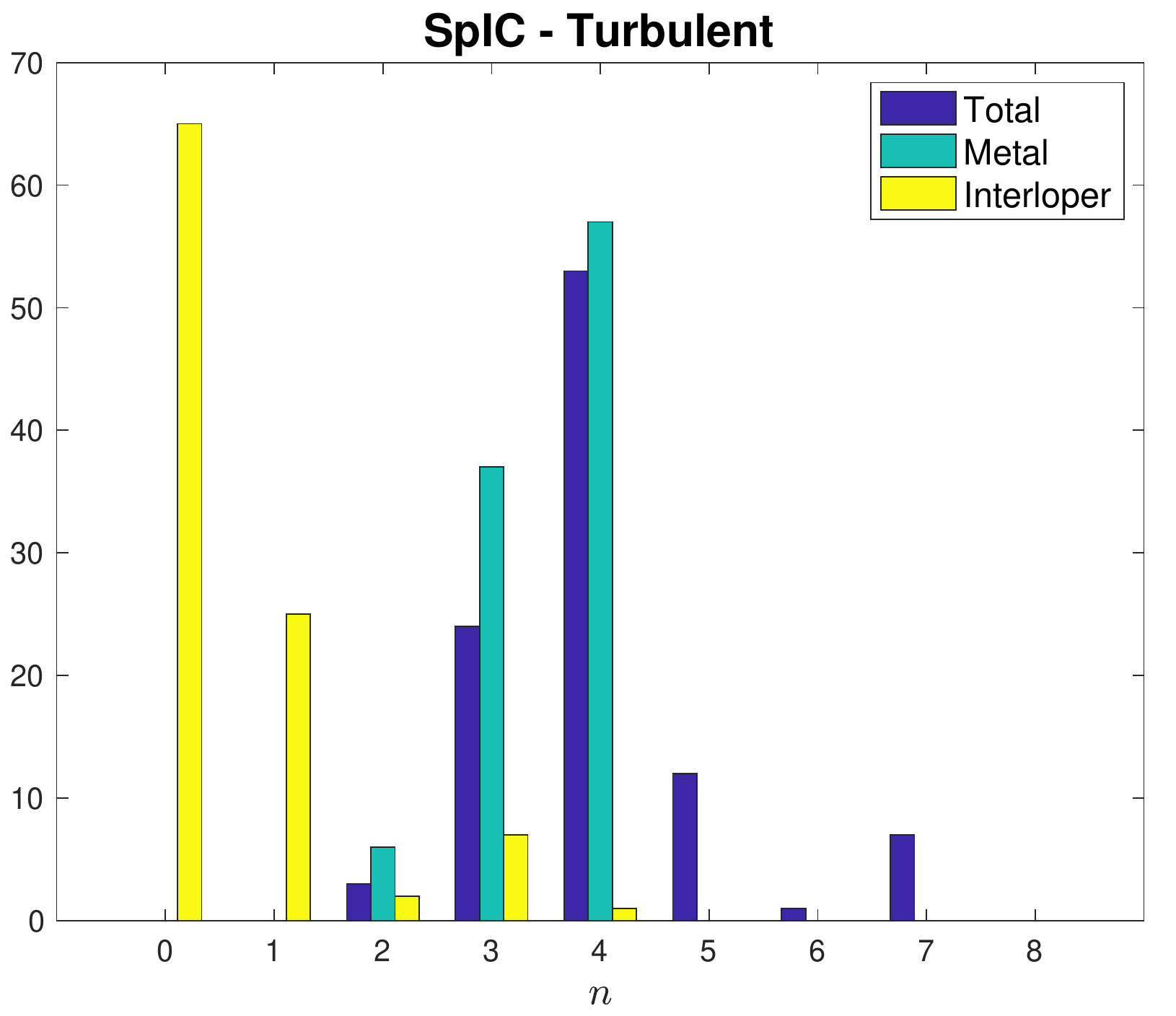}
\includegraphics[width=0.48\linewidth]{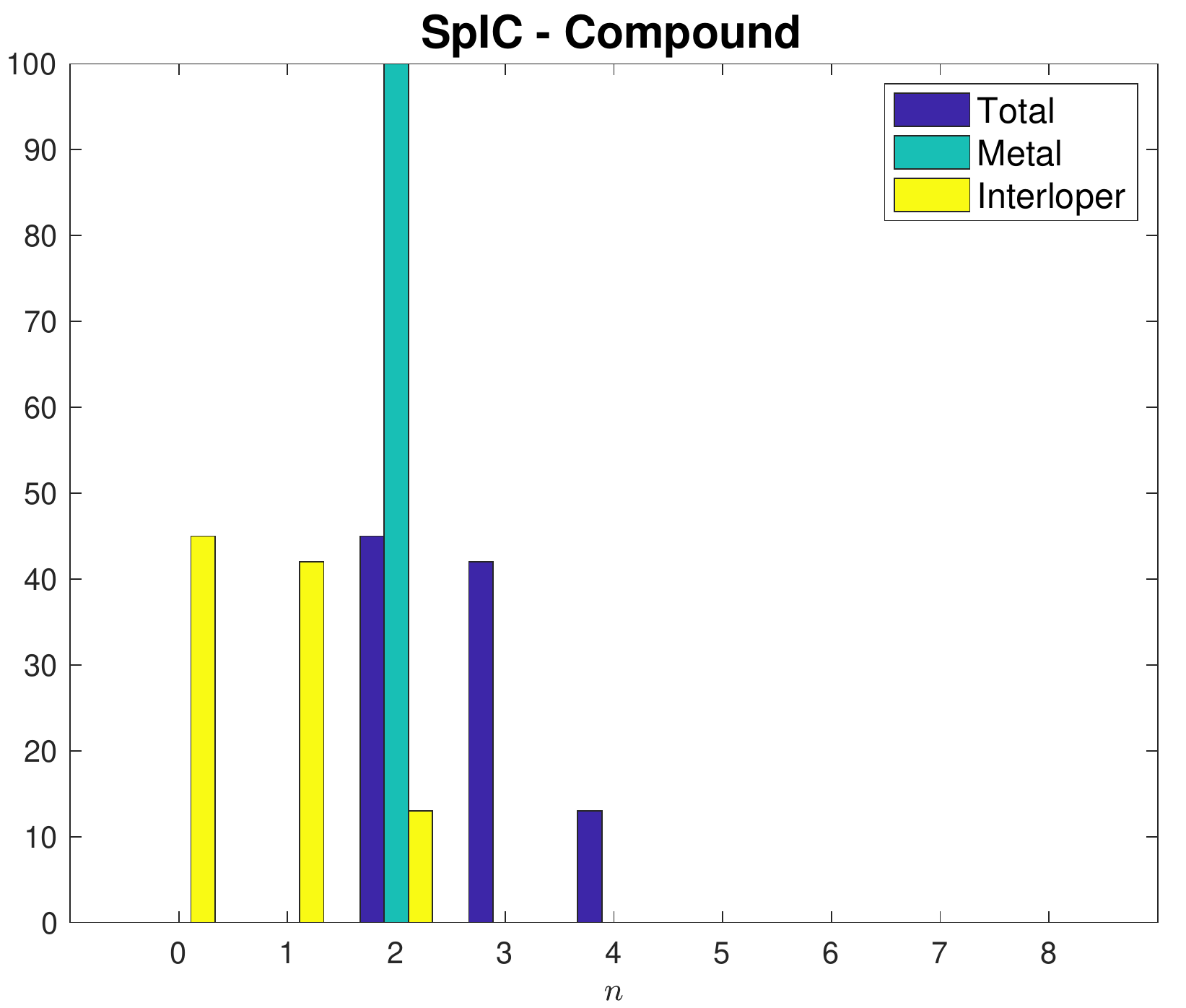} 
\caption{Number of absorption components for the results illustrated in Figure \ref{fig:daa_blinding}. The spectrum being modelled comprised 2 components. Top left: AICc/turbulent results in severely over-fitted data (light blue). Lower left: SpIC/turbulent also over-fits the data but not quite as severely as AICc/turbulent. The turbulent model results have been obtained using distortion blinding, as described in Section \ref{sec:details}. Top right: AICc/compound produces very slight over-fitting. Lower right: SpIC/compound produces no over-fitting. Distortion blinding has {\it not} been used for the compound model results. The yellow bars indicate additional interloper (i.e. components not identified as heavy element) components automatically included in the fits by {\sc ai-vpfit}. These are purely a consequence of chance noise correlations in the input spectrum and appear to have no impact on the final value of $\daa$ measured.
\label{fig:component_blinding}}
\end{figure}

Figure \ref{fig:component_blinding} is a visual breakdown of the number of redshift components needed to fit the spectral lines illustrated in Figure \ref{fig:synthetic_spectrum}, for the four different combinations of AICc, SpIC, turbulent, and compound. Light blue bars correspond to metals (i.e. Mg\,{\sc ii} and Fe\,{\sc ii} in this case). Remembering that the true number of components is 2, we see that the only combination with 100\% success is SpIC/compound, with AICc/compound almost as good. The turbulent results badly overfit the data, both AICc and SpIC giving most probable numbers of 4. The yellow bars correspond to interlopers i.e. additional absorption components required by {\sc ai-vpfit} to achieve an acceptable fit, but which were not identified as metals. No interlopers were present in the simulated data being fitted so these are weak features caused by chance correlated noise patterns that have no detectable impact (in terms of systematic bias) on the $\daa$ estimate.

\begin{figure}
\centering
\includegraphics[width=0.48\linewidth]{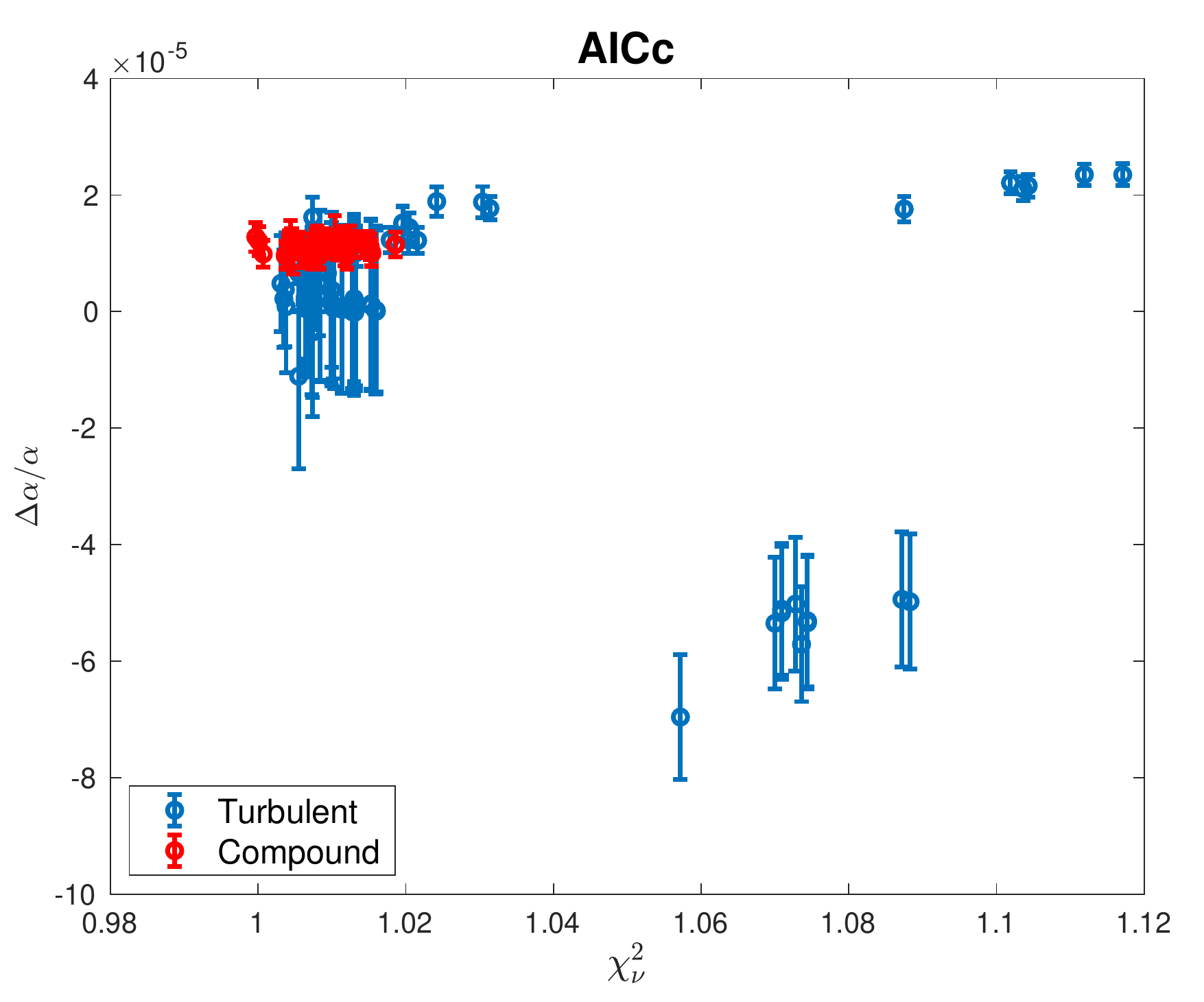}
\includegraphics[width=0.47\linewidth]{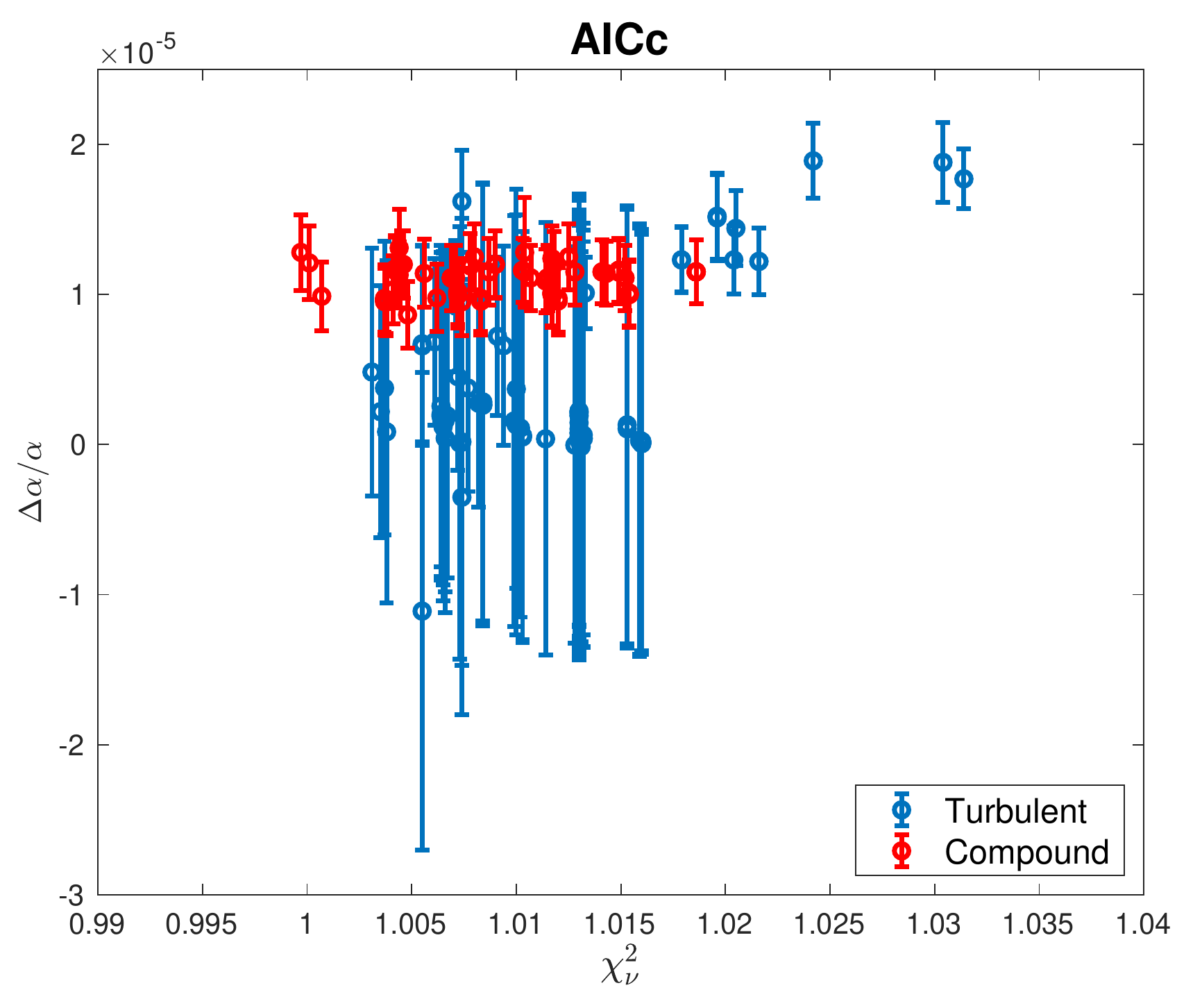} \\
\includegraphics[width=0.48\linewidth]{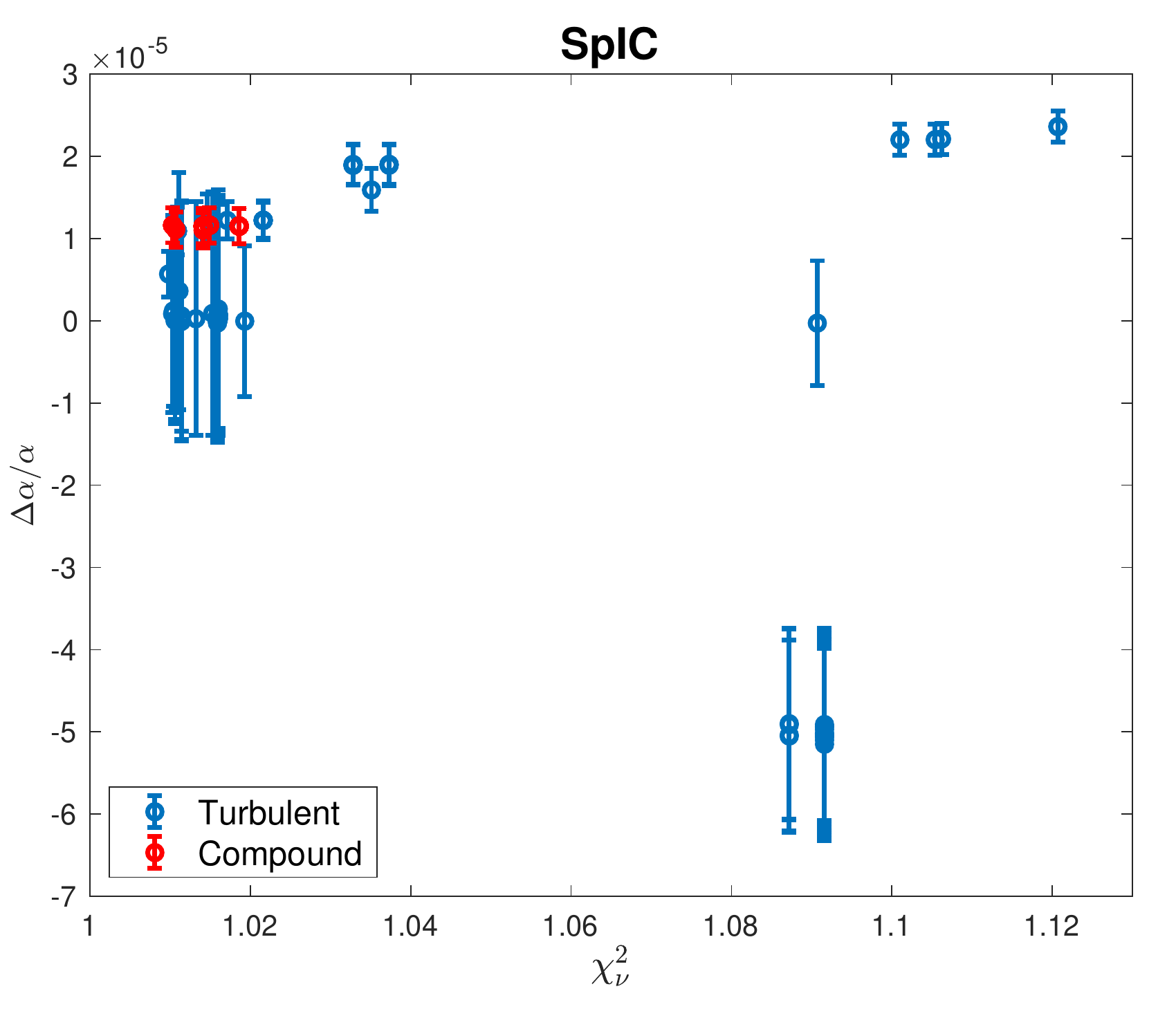}
\includegraphics[width=0.48\linewidth]{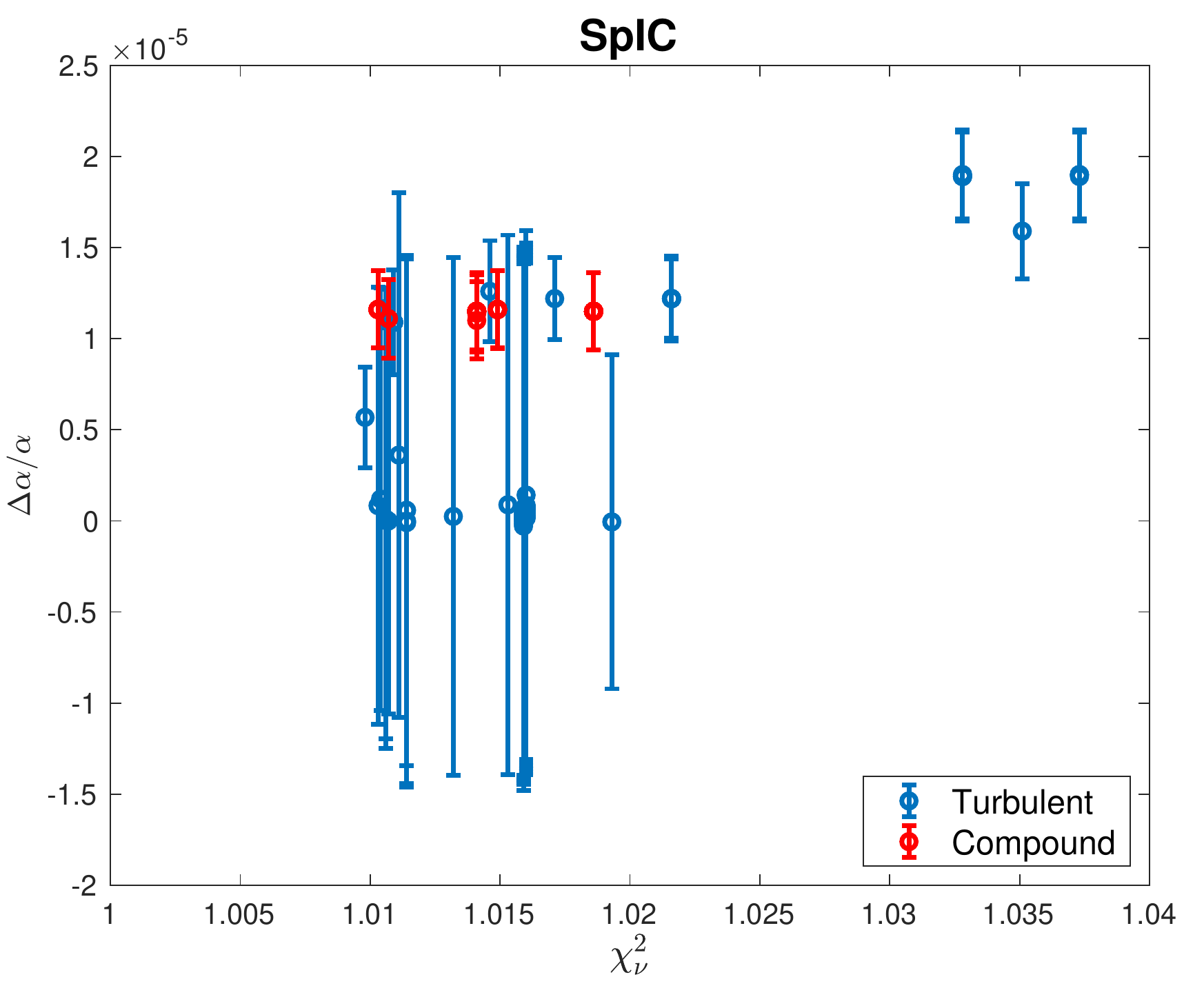} 
\caption{Plots of $\daa$ vs. $\chi_{\nu}^2$ for the 4 combinations of AICc, SpIC, turbulent, and compound. Distortion-blinding was applied to the turbulent (blue) points but not the compound (red) points. See Section \ref{sec:details}. The two right hand panels are zoom-ins of the two left hand panels. \label{fig:chisq-daa_blinding}}
\end{figure}

Figure \ref{fig:chisq-daa_blinding} plots the 100 individual {\sc ai-vpfit} $\daa$ measurements. The distortion-blinded turbulent points (blue) are clearly heavily biased towards zero. Further, the majority of those points have far larger error bars (because of the overfitting and hence artificially increased blending) than a small subset closer to the correct value of $\daa = 8.08 \times 10^{-6}$. The compound points (not distortion-blinded, red) have substantially smaller error bars (no overfitting) and do not suffer bias. Interestingly, the blue points clearly exhibit model non-uniqueness whereas the red ones do not.

\subsection{A retrospective on distortion-blinding}

Distortion-blinding was an attempt at taking a conservative approach to the subject. It is only because we have developed AI Monte-Carlo methods that we can now identify the adverse impact of interactive modellers (each probably taking a slightly different approach) using distortion-blinding + turbulent broadening. {\sc ai-vpfit} shows us how to avoid bias in measurements using high quality data from facilities like ESPRESSO on the VLT and, in the future, HIRES on the ELT. In fact, in the absence of distortion-blinding, any reasonable analysis is effectively inherently blind anyway. If the fitting is done ``manually'' (i.e. using {\sc vpfit}), the process contains so many steps that it is inconceivable for a human to subconsciously construct a biased model unless the final measured value of $\daa$ is influenced directly by iteratively adjusting the model or adding new absorption components until a ``preferred'' value of $\daa$ is obtained. Given that analyses involving distortion-blinding have been carried out by different people at different times, the degree to which this has or has not happened in previously published measurements cannot be recovered. 

Looking to future measurements, no tampering with the wavelength scale needs to be or should be done. If a fit is done using {\sc vpfit}, a far simpler (and {\it unbiased}) method of ``blinding'' would be to switch off all output to the user containing any information about $\daa$ during the entire model construction process. $\daa$ must be a free parameter from the outset i.e. as soon as two or more transitions or species with differing sensitivities to $\alpha$ are incorporated -- see \cite{Lee2020AI-VPFIT} for a discussion on this important point. Switching off $\daa$ output naturally renders any conscious or subconscious steering by the user impossible. 

Analyses using the {\sc ai-vpfit} method, as described in \cite{Lee2020AI-VPFIT}, are automatically blinded. There is no interactive input during the modelling procedure.

\section{Discussion} \label{sec:howto}

The calculations described in this work and accompanying discussions allow us to define some general requirements for spacetime measurements of varying fundamental constants:

\begin{enumerate}
\item When modelling/solving for $\daa$, if fitting multiple species simultaneously, turbulent broadening should not be used. Instead only compound broadening should be used. If compound broadening is problematic because $T$ is not sufficiently well constrained, either the measurement should be discarded, or possibly one could adopt a representative T as a fixed parameter, although in this case the uncertainty on $\daa$ may be artificially lowered. We have not investigated such an approach in this paper.
\item Distortion-blinding biases results and should be avoided. Deliberately distorting a spectrum and then solving for its velocity structure with $\daa$ forced to be zero is most likely to create a result that is biased towards $\daa=0$. Previous $\daa$ measurements carried out in this way should be repeated.
\item Some absorption systems suffer from model non-uniqueness. Therefore it is desirable to model each absorption system multiple times to quantify this.
\item In obtaining the observational data, wavelength coverage should be done using LFCs or FPs. If observations from different observing runs at different epochs are to be combined/jointly analysed, there should be calibration redundancy i.e. ideally calibrations always done with two LFCs or FPs. Accurate calibration reaching as blue as the atmospheric cutoff is essential to pick up lower rest-wavelength transitions.
\end{enumerate}

An aside on the above is an interesting consequence of turbulent broadening with important implications for abundance measurements. Turbulent modelling generates spurious absorption components and these appear with meaningless relative abundances. The result is that the relative abundances from one absorption component to another appear to be highly (artificially) scattered. This is seen in Section \ref{sec:preliminary} and Figures \ref{fig:exaggerated_compound} and \ref{fig:exaggerated_turbulent}. Let $\Delta = \log N$(Mg\,{\sc ii})$ - \log N$(Fe\,{\sc ii}). Table \ref{tab:single} gives $\Delta$(true) $=0.5$, $\Delta$(compound) $=0.5$, $\Delta$(turbulent, $v=-2.152$) $=-1.3$, $\Delta$(turbulent, $v=-0.279$) $\sim 2.5$, where, for the purposes of this illustration, the latter assumes a column density detection threshold of 10.0, so is an order of magnitude estimate. This means that claims for significant relative abundance fluctuations across an absorption complex may be incorrect if turbulent broadening has been assumed. If, instead, {\it summed} column densities are used for a complex, no spurious effects would be seen.

\vspace{6pt} 
\authorcontributions{All authors contributed equally to this work.}

\funding{We are grateful to the John Templeton Foundation for support.} 

\dataavailability{We thank the ESPRESSO consortium for making public the reduced ESPRESSO spectrum of HE0515--4414 that we used for some of the calculations in this paper.}

\acknowledgments{The supercomputer calculations described in this paper were carried out using OzSTAR at the Centre for Astrophysics and Supercomputing at Swinburne University of Technology. We also thank Mariusz P. Dąbrowski and Vincenzo Salzano for organising the Alternative Gravities and Fundamental Cosmology ALTECOSMOFUN'21 conference, Szczecin, Poland, 6-10 September 2021, and Bob Carswell and Paolo Molaro for comments on a draft of this paper.}

\conflictsofinterest{The authors declare no conflict of interest.}

\reftitle{References}
\externalbibliography{yes}
\bibliography{howtobib}

\end{document}